\newcommand{\CLASSINPUTtoptextmargin}{50pt}
\newcommand{\CLASSINPUTbottomtextmargin}{50pt}
\newcommand{\CLASSINPUTinnersidemargin}{41pt}
\newcommand{\CLASSINPUToutersidemargin}{41pt}

\documentclass[9pt, sigconf,screen=true,bookmarks=false]{acmart}

\usepackage[utf8]{inputenc} %
\usepackage[T1]{fontenc}    %
\usepackage{xcolor}         %

\usepackage{algorithm}
\usepackage{graphicx}
\usepackage{threeparttable}
\usepackage{multirow}
\usepackage{amsmath}
\usepackage{bm}
\usepackage{bbm}
\usepackage{amsthm}
\usepackage[]{algpseudocode}
\usepackage{enumitem} %
\usepackage[subrefformat=parens,labelformat=parens]{subfig}

\usepackage{wrapfig}
\usepackage[algo2e, ruled, vlined, linesnumbered, noend]{algorithm2e}

\definecolor{citecolor}{RGB}{34,139,34}
\definecolor{mydarkblue}{rgb}{0,0.08,1}
\definecolor{mydarkgreen}{rgb}{0.02,0.6,0.02}
\definecolor{mydarkred}{rgb}{0.8,0.02,0.02}
\definecolor{mydarkorange}{rgb}{0.40,0.2,0.02}
\definecolor{mypurple}{RGB}{111,0,255}
\definecolor{myred}{rgb}{1.0,0.0,0.0}
\definecolor{mygold}{rgb}{0.75,0.6,0.12}
\definecolor{myblue}{rgb}{0,0.2,0.8}
\definecolor{mydarkgray}{rgb}{0.,0.2,0.2}

\definecolor{lightred}{RGB}{255,235,235}
\definecolor{lightgreen}{RGB}{235,255,235}
\definecolor{lightblue}{RGB}{235,235,255}
\definecolor{lightcyan}{RGB}{235,255,255}
\definecolor{lightmagenta}{RGB}{255,235,255}
\definecolor{lightyellow}{RGB}{255,255,235}

\definecolor{qxkcolor}{RGB}{215,235,255}
\definecolor{softmaxcolor}{RGB}{230,235,255}
\definecolor{probxvcolor}{RGB}{255,255,235}

\definecolor{topkcolor}{RGB}{255,235,235}
\definecolor{zecolor}{RGB}{255,255,235}
\definecolor{dynacolor}{RGB}{235,255,255}

\definecolor{reviewcolor}{RGB}{0,0,200}

\newcommand{\floor}[1]{\lfloor #1 \rfloor}
\newcommand{\ceil}[1]{\lceil #1 \rceil}

\newcommand{\calA}{\mathcal{A}}
\newcommand{\calD}{\mathcal{D}}
\newcommand{\calO}{\mathcal{O}}
\newcommand{\calF}{\mathcal{F}}

\newcommand{\calN}{\mathcal{N}}
\newcommand{\calL}{\mathcal{L}}
\newcommand{\calQ}{\mathcal{Q}}
\newcommand{\calP}{\mathcal{P}}
\newcommand{\calR}{\mathcal{R}}

\newcommand{\calT}{\mathcal{T}}

\DeclareMathOperator*{\argmin}{argmin}

\theoremstyle{plain}

\theoremstyle{definition}

\algdef{SE}[DOWHILE]{Do}{doWhile}{\algorithmicdo}[1]{\algorithmicwhile\ #1}%

\newcommand{\squishlist}{
 \begin{list}{$\bullet$}
  { \setlength{\itemsep}{0pt}
     \setlength{\parsep}{3pt}
     \setlength{\topsep}{3pt}
     \setlength{\partopsep}{0pt}
     \setlength{\leftmargin}{1.5em}
     \setlength{\labelwidth}{1em}
     \setlength{\labelsep}{0.5em} } }

\newcommand{\squishend}{
  \end{list}  }

\newcommand{\dc}{\texttt{DC}\xspace}
\newcommand{\ps}{\texttt{PS}\xspace}
\newcommand{\crs}{\texttt{CR}\xspace}
\newcommand{\supermesh}{\texttt{SuperMesh}\xspace}
\newcommand{\submesh}{\texttt{SubMesh}\xspace}

\graphicspath{{./figs/}}
\newcommand{\name}{\texttt{ADEPT}\xspace}
\usepackage{geometry}
\copyrightyear{2022}
\acmYear{2022}
\setcopyright{acmcopyright}\acmConference[DAC '22]{Proceedings of the 59th ACM/IEEE Design Automation Conference (DAC)}{July 10--14, 2022}{San Francisco, CA, USA}
\acmBooktitle{Proceedings of the 59th ACM/IEEE Design Automation Conference (DAC) (DAC '22), July 10--14, 2022, San Francisco, CA, USA}
\acmPrice{15.00}
\acmDOI{10.1145/3489517.3530562}
\acmISBN{978-1-4503-9142-9/22/07}

\begin{document}
\settopmatter{printacmref=false} %

\pagestyle{plain} %

\title{
APT: \underline{A}utomatic \underline{P}hotonic \underline{T}ensor Core Design via Hardware-Adaptive Circuit Exploration
}
\title{
ADEPT: \underline{A}utomatic \underline{D}ifferentiable D\underline{E}sign of\\ \underline{P}hotonic \underline{T}ensor Cores
}

\author
{
Jiaqi Gu,
Hanqing Zhu,
Chenghao Feng,
Zixuan Jiang,
Mingjie Liu,
Shuhan Zhang,\\
Ray T. Chen,
David Z. Pan\\
University of Texas at Austin\\
\small\textit{\{ jqgu, hqzhu, fengchenghao1996, zixuan, jay\_liu, shuhan.zhang \}@utexas.edu,~\{ chen, dpan \}@ece.utexas.edu}
}
\begin{abstract}
\label{abstract}
Photonic tensor cores (PTCs) are essential building blocks for optical artificial intelligence (AI) accelerators based on programmable photonic integrated circuits.
PTCs can achieve ultra-fast and efficient tensor operations for neural network (NN) acceleration.
Current PTC designs are either manually constructed or based on matrix decomposition theory, which lacks the adaptability to meet various hardware constraints and device specifications.
To our best knowledge, automatic PTC design methodology is still unexplored.
It will be promising to move beyond the manual design paradigm and "nurture" photonic neurocomputing with AI and design automation.
Therefore, in this work, for the first time, we propose a fully differentiable framework, dubbed \name, that can efficiently search PTC designs adaptive to various circuit footprint constraints and foundry PDKs.
Extensive experiments show superior flexibility and effectiveness of the proposed \name framework to explore a large PTC design space.
On various NN models and benchmarks, our searched PTC topology outperforms prior manually-designed structures with competitive matrix representability, \textbf{2$\times$-30$\times$} higher footprint compactness, and better noise robustness, demonstrating a new paradigm in photonic neural chip design.
The code of \name is available at \href{https://github.com/JeremieMelo/ADEPT}{link} using the \href{https://github.com/JeremieMelo/pytorch-onn}{TorchONN} library.
\end{abstract}

\maketitle

\section{Introduction}
\label{sec:Introduction}
With the advance in integrated photonics, the optical neural network (ONN) has become a promising candidate for ultra-efficient deep neural network (DNN) acceleration~\cite{NP_NATURE2017_Shen,NP_APR2020_Miscuglio, NP_PIEEE2020_Cheng, NP_NaturePhotonics2021_Shastri}.
As light propagates through the photonic integrated circuits (PICs), computation-intensive matrix multiplication can be achieved with sub-nanosecond latency and near-zero energy consumption~\cite{NP_APR2020_Miscuglio}.
Shen \textit{et al.}~\cite{NP_NATURE2017_Shen} demonstrated a triangular photonic mesh with cascaded Mach-Zehnder interferometers (MZIs) to realize matrix multiplication using optics.
They use singular value decomposition (SVD) to decompose the weight matrix $W$ into $U\Sigma V$, parametrize the unitary matrices $U$ and $V$ with a series of planar rotators, and map them into a triangular photonic mesh.
This matrix-decomposition-based photonic tensor core (PTC) design is universal but suffers from high area cost and unsatisfying noise robustness.
To improve the area efficiency, a Fourier-transform (FFT) based PTC~\cite{NP_ASPDAC2020_Gu, NP_TCAD2020_Gu, NP_Arxiv2021_Feng} was proposed that shrinks the circuit depth from linear to logarithmic using a butterfly circuit topology.
This design removes large MZIs and constructs the PTC with smaller basic optical components instead.

However, previous PTCs are all hand-designed based on matrix decomposition theory, which leaves a large design space unexplored and lacks the adaptability to meet various device specifications and hardware constraints.
Specifically, the MZI-based PTC~\cite{NP_NATURE2017_Shen} is universal at the cost of high area cost and low compute density.
The FFT-based PTC~\cite{NP_ASPDAC2020_Gu, NP_TCAD2020_Gu, NP_Arxiv2021_Feng} significantly reduces the usage of couplers and phase shifters.
However, its area efficiency may not scale well with different PTC sizes and foundry process design kits (PDKs).
As the PTC size scales up, the butterfly mesh in the FFT-based PTC introduces quadratically many waveguide crossings.
If the foundry does not provide compact crossings, e.g., AIM photonics~\cite{NP_OFC2018_Timurdogan},
those routing-related crossings will take up most of the circuit area.
Besides, the butterfly mesh only has a logarithmic depth.
Thus it restricts the matrix representability, which may lead to inadequate ONN learnability as PTC scales up.

Based on the above analysis, we observe strong demand for an automatic, efficient, and flexible PTC design methodology.
Inspired by the success of neural architecture search (NAS)~\cite{NN_ICLR2018_Liu, NN_CVPR2019_Wu} in the machine learning community, an interesting question to be answered is \emph{whether we can jump out of the conventional manual design paradigm and use AI to "nurture" photonic neurocomputing with higher flexibility}.
However, PTC design search encounters the following unique and difficult challenges.
First, unlike NAS, where the NN architecture can be re-designed in software for application/platform adaptation at a relatively low cost, the photonic circuits need to be carefully designed before chip manufacturing and cannot be easily changed given the \emph{high cost of chip tape-out}.
Second, the PTC design can only be searched on a proxy NN model and learning task, but it has to be expressive and general enough to be \emph{adapted to various AI workloads} after chip manufacturing.
Third, the PTC circuit topology has an extremely large and highly discrete search space, which casts \emph{significant optimization difficulties} that prevents direct application of off-the-shelf NAS algorithms to this unique problem.

To handle those challenges, in this work, we propose the \emph{first} automatic differentiable search framework for photonic tensor core topology design, which we refer to as \name.
Our target is, given certain footprint constraints, we can efficiently search for a photonic circuit topology with good matrix \emph{representability}, compact \emph{footprint}, and high noise \emph{robustness}.
\name enables differentiable PTC topology exploration via the following approaches:
(1) we construct a probabilistic photonic \supermesh to enable differentiable optimization in a huge and highly discrete PTC search space;
(2) we adopt reparametrization and augmented Lagrangian method to learn waveguide connections;
(3) binarization-aware training is employed to learn the location to place optical couplers;
(4) \name integrates the device specification from foundry PDKs into the \supermesh training flow and optimizes PTC designs under various footprint constraints in a fully differentiable approach.

Our main contributions are as follows,
\squishlist
    {\item In this work, \emph{for the first time}, we automate the photonic tensor core design process and propose a differentiable framework to efficiently explore the PTC design space.}
    {\item To enable PTC topology search in a differentiable way, we introduce probabilistic photonic \supermesh to search the PTC depth, augmented Lagrangian method to learn waveguide connections, and binarization-aware training to learn the coupler placement.}
    {\item The proposed \name flow can adaptively generate various PTC designs based on different foundry PDKs and circuit footprint constraints.
    Experiments on various NN models and datasets show that the searched PTC topology outperforms prior hand-crafted structures with higher flexibility, competitive expressiveness, \textbf{2$\times$-30$\times$} smaller footprint, and superior noise robustness.}
\squishend

\section{Background}
\label{sec:Background}
\subsection{Photonic Computing Basics}
\label{sec:PhotonicsBasics}
To perform neurocomputing in optics, we construct photonic integrated circuits (PICs) by cascaded optical devices.

\noindent\textbf{Phase shifter (\ps).}~
Phase shifters can manipulate the effective refractive index of waveguides to produce a controlled phase shift $\phi$ on the propagating light signal $x$, $y=e^{-j\phi}x$.
Phase shifters are typically active devices that are reprogrammable after PIC manufacturing.

\noindent\textbf{Directional coupler (\dc).}~
2-by-2 directional couplers (\dc) can produce interference between two coherent light signals, whose transfer matrix is $T_{2\times 2}$, where $T_{11}=T_{22}=t$ and $T_{12}=T_{21}=\sqrt{1-t^2}j$, and $t\in[0,1]$ is the transmission coefficient.
Couplers are typically passive devices that are fixed after chip fabrication.

\noindent\textbf{Waveguide Crossing (\crs).}~
Given the 2-dimensional topology of the PIC, signal routing requires waveguide crossings.
Unlike electrical wires, silicon waveguides allow independent light propagation through crossed waveguides.
Crossings of $n$ waveguides can be described as an $n\times n$ permutation matrix.
In photonic tensor cores, crossings can enhance signal flow and are typically not programmable after PIC fabrication.

\noindent\textbf{Mach-Zehnder interferometer (MZI).}~
MZI is a hand-designed structure consisting of two cascaded couplers and two phase shifters.
MZI can perform arbitrary 2-D unitary projection, which is widely used to construct PTCs at the cost of a large circuit footprint.

\subsection{Programmable PTCs}
\label{sec:ProgrammablePTC}
PTCs are essential building blocks in photonic accelerators constructed with passive and active optical devices.
Current PTC topologies are hand-designed and barely involve any automation.
Various~\cite{NP_NATURE2017_Shen,NP_Optica2018_Clements} MZI meshes were proposed to realize arbitrary $N\times N$ unitary matrices using $N(N-1)/2$ cascaded MZIs.
Based on this, a weight matrix can be decomposed using SVD and mapped onto MZI meshes.
Besides this universal photonic mesh design, a Fourier transform (FFT)-based PTC design~\cite{NP_ASPDAC2020_Gu, NP_TCAD2020_Gu} was introduced to realize restricted linear operations with a butterfly-style mesh topology.
This design utilizes basic optical components, i.e., \ps, \dc, and waveguide crossings \crs without large MZIs to reduce footprint.

PTC designs need to consider device specification in foundry PDKs to honor circuit footprint constraints.
Different foundries, e.g., AMF~\cite{NP_AMF} and AIM photonics~\cite{NP_OFC2018_Timurdogan}, provide devices of considerably different sizes, which makes it challenging to manually search for good PTC designs that fit the area budget.
This motivates us to provide an automatic solution for PDK-adaptive PTC design.

\subsection{Differentiable Neural Architecture Search}
\label{sec:DNAS}
Differentiable neural architecture search (DNAS) is widely adopted to automate the manual process of DNN architecture design with high
efficiency.
DNAS relaxes the discrete search space into continuous representation, such that the architecture can be optimized with gradient-based methods.
DARTS~\cite{NN_ICLR2018_Liu} enables DNAS by using a softmax function to relax the categorical choice of candidate operations.
FBNet~\cite{NN_CVPR2019_Wu} represents the search space by a stochastic SuperNet and then applies DNAS to discover low-latency DNN designs.

Recently, O-HAS~\cite{NP_ICCAD2021_Li} proposed an optical accelerator search framework that can automatically generate the optimal accelerator architecture.
Different from our PTC circuit topology design, O-HAS focuses on searching for a mapping strategy to implement DNN models with manually-designed PTCs.

To the best of our knowledge, automated PTC design flow remains unexplored.
It will be promising to develop a flexible and efficient framework to automatically search PTC topologies with high expressiveness, compact footprint, and good noise robustness, adaptive to various PDKs and footprint constraints.  

\section{Automatic Photonic Tensor Core Design Framework \name}
\label{sec:Method}
\begin{figure}
    \centering
    \includegraphics[width=0.9\columnwidth]{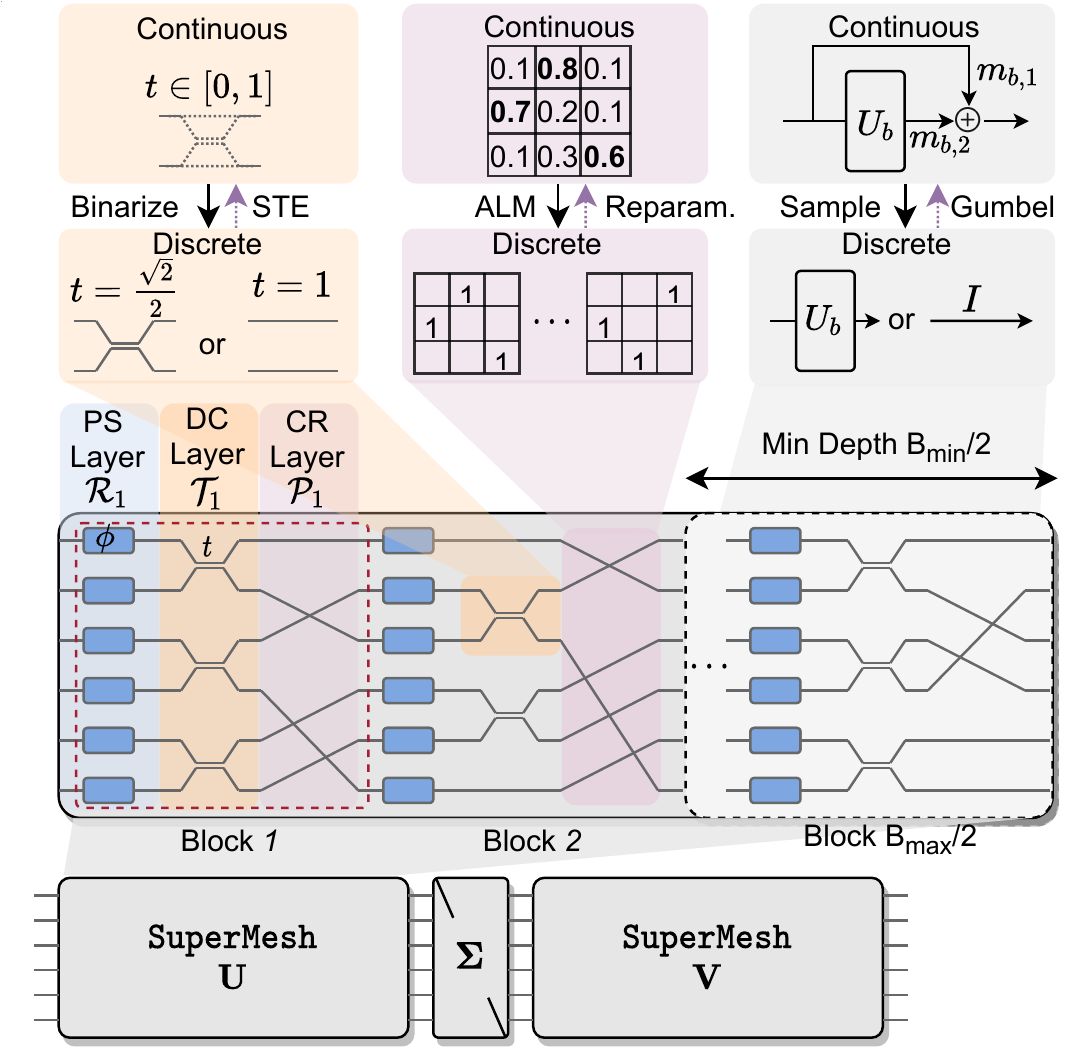}
    \vspace{-10pt}
    \caption{Overview of the probabilistic photonic \supermesh.}
    \label{fig:Overview}
    \vspace{-5pt}
\end{figure}

\subsection{Problem Formulation}
Our target is to use basic optical components, including \dc, \ps, and \crs, to design a photonic mesh with a controlled footprint that can construct ONNs with high expressiveness, formulated as follows,
\begin{equation}
\label{eq:Formulation}
\begin{aligned}
&\min_{\alpha\in\calA} \calL\big(W^{*\alpha};~\calD^{val}\big), \quad \alpha=(B^U, B^V, \calP, \calT)\\
\text{s.t.} ~~&W^* = \argmin_{W} \calL ( W^{\alpha}; ~ \calD^{trn}),~~ F_{min}\leq \calF(\alpha) \leq F_{max},\\
&W^{\alpha}\in{\mathbb{C}^{M\times N}}=\big\{W^{\alpha}_{pq}\big\}_{p=1,q=1}^{p=P,q=Q}=\big\{U^{\alpha}_{pq}\Sigma_{pq}V_{pq}^{\alpha}\big\}_{p=1,q=1}^{p=P,q=Q},\\
&B^U,B^V\in[B_{min}/2,B_{max}/2], W_{pq}\in\mathbb{C}^{K\times K},\\ 
&\calP\!=\!(\cdots,\calP_b,\cdots,\calP_{B^U+B^V}),\calT\!=\!(\cdots,\calT_b,\cdots,\calT_{B^U+B^V}).
\end{aligned}    
\end{equation}
The weight matrix $W$ in an ONN layer is partitioned into $K \times K$ sub-matrices.
Each sub-matrix is constructed by two unitaries $U_{pq}^{\alpha}$ and $V_{pq}^{\alpha}$ and a diagonal matrix $\Sigma_{pq}$.
The layout topology $\alpha$ of two unitaries is the primary search target, shared among all blocks.

\subsection{Search Space Specification}
\label{sec:SearchSpaceSpec}
Our search space focuses on the \emph{tensor core circuit topology}, not layer configurations like conventional NAS work.
As illustrated in Fig.~\ref{fig:Overview}, we define the following block-wise search space for the unitaries,
\begin{equation}
    \label{eq:UnitarySearchSpace}
    U^{\alpha}_{pq}=\prod_{b=1}^{B^U}\calP_b\calT_b\calR(\Phi_{pq}^{b}),\quad V^{\alpha}_{pq}=\prod_{b=B^U+1}^{B^U+B^V}\calP_b\calT_b\calR(\Phi_{pq}^{b}).
\end{equation}
Unitaries $U$ and $V$ consist of $B^U$ and $B^V$ blocks, respectively. 
For simplification, we focus on $U$ and refer to $\{B^U,B^V\}$ as $B$ thereafter.

The first structure in the block is a column of $K$ phase shifters, which can be described by a diagonal matrix $\calR(\Phi_{pq}^b)$,
\begin{equation}
    \label{eq:PSLayer}
    \calR(\Phi_{pq}^b)=\texttt{diag}(e^{-j\phi_1},\cdots,e^{-j\phi_K}).
\end{equation}

The second structure in the block is a column of 2-by-2 directional couplers $T$'s placed from the $s_b$-th waveguide, which can be described by a block diagonal matrix $\calT_b$.
We will only include 50:50 DCs in our design, i.e., $t=\sqrt{2}/2$.
This coupler column enables information interaction between adjacent waveguides.
Besides, cascading \dc layers in an \emph{interleaved} way naturally allows more light signals to interfere with each other.
Thus we have $s_b=1$ if $b$ is even and $s_b=0$ if $b$ is odd, as shown in Fig.~\ref{fig:Overview}.

The last layer in the block is designed for pure waveguide routing.
This layer consists of a network of waveguide crossings, whose transfer matrix belongs to the permutation matrix family,
\begin{equation}
\small
    \label{eq:PermutationLayer}
    \begin{aligned}
    &\calP_b\in\!\{0,1\}^{K\times K}, \sum_j\calP_b^{i,j}\!=\!1~ \forall i\in\![K],
    \sum_i\calP_b^{i,j}\!=\!1~\forall j\in\![K].
    \end{aligned}
\end{equation}
The search space for $\calP$ is extremely large since $B_{max}$ size-$K$ permutations contain total $(K!)^{B_{max}}$ possible combinations.

In summary, one unitary photonic mesh contains $B$ blocks, each including a \ps layer, a \dc layer, and a \crs layer.
The topology $\alpha$ includes the number of blocks $B^U$ and $B^V$, the waveguide connections $\calP$ in the permutation layer, and the locations to put directional couplers described by $\calT$.
The total search space is $\calO\big((K\cdot K!/2)^{B_{max}}\big)$.

\subsection{Fully Differentiable \texttt{SuperMesh} Training}
\begin{figure}
    \centering
    \includegraphics[width=\columnwidth]{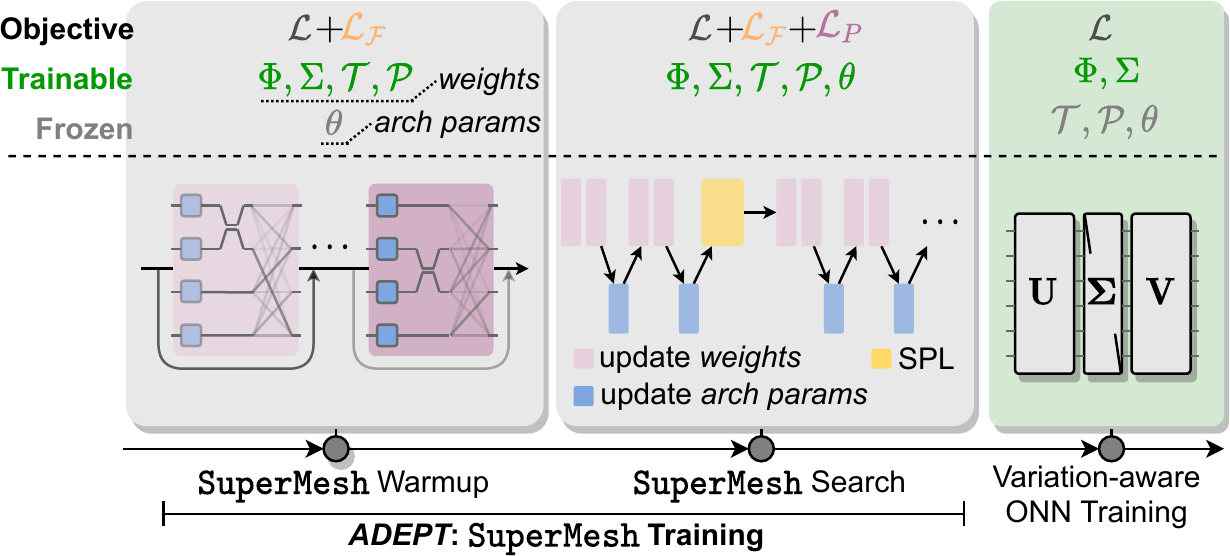}
    \vspace{-10pt}
    \caption{The proposed photonic \supermesh training flow \name, followed by variation-aware ONN training.}
    \label{fig:TrainFlow}
    \vspace{-5pt}
\end{figure}

To solve the highly discrete PTC topology design problem in such an enormous search space, we propose a differentiable \supermesh training flow \name in Fig.~\ref{fig:TrainFlow}.

The total optimization variables in the \supermesh training contain (1) diagonal matrix $\Sigma$, (2) phases $\Phi$ in the \ps layer, (3) directional couplers $\calT$ in the \dc layer, (4) permutation matrices $\calP$ of the \crs layer, (5) the number of blocks $B$. 
Jointly optimizing all those continuous and discrete variables is highly ill-conditioned, leading to prohibitive optimization difficulty.
We separate them into two sets: (1) $\Sigma$, $\Phi$, $\calT$, and $\calP$ belong to the \supermesh \emph{weights}, and (2) $B$ belong to the \emph{architecture parameter} group.
The entire \name flow contains two stage, shown in Fig.~\ref{fig:TrainFlow}.
The first \supermesh \emph{Warmup} stage only optimizes \emph{weights} for initial exploration.
The second \supermesh \emph{Search} stage optimizes two parameter groups alternately.
We periodically enter the \supermesh \emph{weight training} phase to optimize $\Sigma$, $\Phi$, $\calT$, and $\calP$ and
switch to the \emph{architecture parameter training} phase to search $B$.
After \name \supermesh training, we apply variation-aware training to target ONN models with the searched PTC topology.
Now we introduce how to optimize those variables one by one.

\subsubsection{Optimize \supermesh Depth $B$}~
\label{sec:OptimizeBlock}
The depth of \supermesh can be relaxed by constructing a stochastic super block.
During the inference, the $b$-th block $U_b$ is either sampled and executed ($U_{b,1}$) or skipped as an identity projection ($U_{b,2}$) with the probability of
\begin{equation}
    \small
    \label{eq:SuperBlockProb}
    P_{\theta_b}\big(U_b=U_{b,i}\big)=e^{\theta_{b,i}}\Big/\sum_{i}e^{\theta_{b,i}}.
\end{equation}
The probability distribution of block-$b$ is parametrized by the sampling coefficient $\theta_b$.
The forward propagation of the $b$-th block is,
\begin{equation}
    \small
    \label{eq:SuperBlockForward}
    x_{b+1}=\sum_{i=1}^2m_{b,i}U_{b,i}x_{b}, \quad U_{b,1}=I,~U_{b,2}=\calP_b\calT_b\calR_{b},
\end{equation}
where the variable $m_{b,i}$ determines the probability to select the $b$-th block.
Therefore, instead of searching $B$ in the discrete space, the problem can be relaxed to the optimization of the probability $P_{\theta}$.
Gumbel-Softmax (GS) trick~\cite{NN_CVPR2019_Wu} is employed as follows,
\begin{equation}
    \small
    \label{eq:GumbelSoftmax}
    m_{b,i}=\texttt{GumbelSoftmax}(\theta_{b,i}|\theta_b)=e^{(\theta_{b,i}+g_{b,i})/\tau}\Big/\sum_{i}e^{(\theta_{b,i}+g_{b,i})/\tau}.
\end{equation}
$\texttt{Softmax}$ achieves continuous relaxation, and the Gumbel noise $g_{b,i}$ introduces stochasticity for better exploration controlled by the temperature $\tau$.
Note that the depth $B$ has a range of $[B_{min}/2,B_{max}/2]$.
Hence, the \supermesh $U$ consists of $(B_{max}/2)$ super blocks to upper-bound the search space.
Meanwhile, the last $(B_{min}/2)$ blocks are always sampled with 100\% certainty to lower-bound the search space, i.e., $m_{b,2}=1, \forall b> B_{max}/2-B_{min}/2$.

\subsubsection{Optimize Permutation Matrices $\calP$}~
\label{sec:ALMPermutation}
\begin{figure}
    \centering
    \includegraphics[width=\columnwidth]{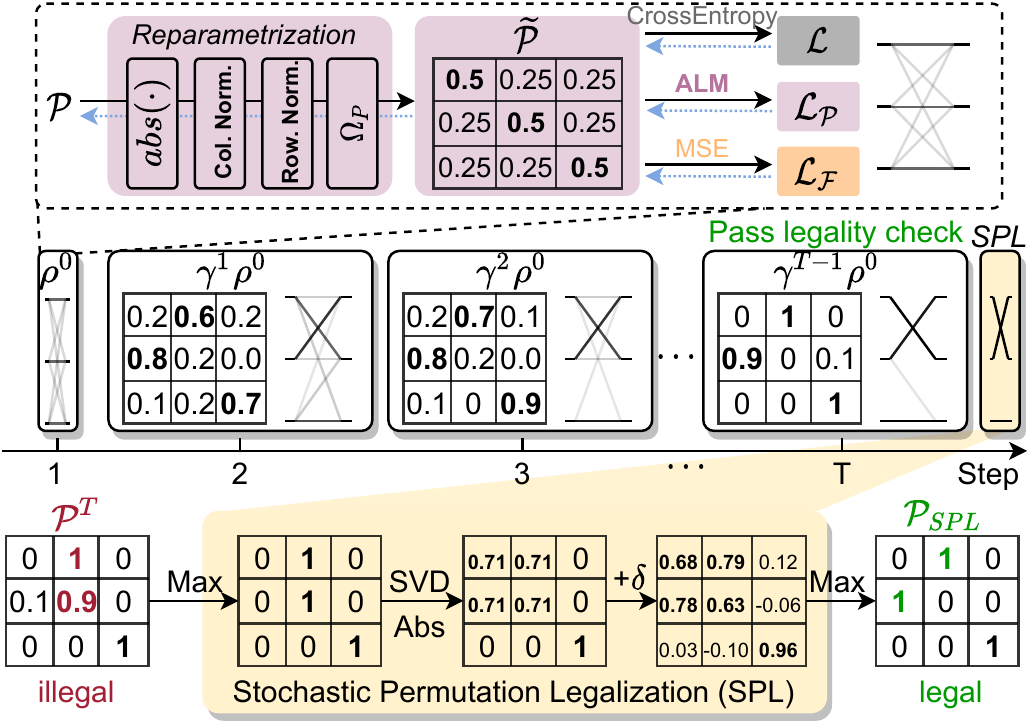}
    \vspace{-15pt}
    \caption{\emph{Top}: permutation optimization procedure.
    \emph{Bottom}: an example for stochastic permutation legalization (\texttt{SPL}).
    }
    \label{fig:PermOptimization}
    \vspace{-5pt}
\end{figure}

Permutations are hard to search directly due to the factorial and highly discrete search space.
The discrete constraint in Eq.~\eqref{eq:PermutationLayer} has a continuous format~\cite{NN_CVPR2017_Cruz},
\begin{equation}
    \label{eq:PermutationLayerConvert}
    \begin{aligned}
    \calP_b\geq 0;~\|\calP_b^{i,:}\|_1=\|\calP_b^{i,:}\|_2, \forall i;~\|\calP_b^{:,j}\|_1=\|\calP_b^{:,j}\|_2, \forall j,
    \end{aligned}
\end{equation}
where the row-wise and column-wise $\ell_1$-norm equals to the $\ell_2$-norm. 
Eq.~\eqref{eq:PermutationLayerConvert} can be relaxed to its convex hull, i.e., Birkhoff polytope,
\begin{equation}
    \label{eq:PermutationLayerRelax}
    \begin{aligned}
    \calP_b\geq 0,~ \textbf{1}^T\calP_b=\textbf{1}^T,~\calP_b\textbf{1}=\textbf{1},~\textbf{1}=(1,\cdots,1)^T.
    \end{aligned}
\end{equation}

As shown in Fig.~\ref{fig:PermOptimization}, we use 1) \emph{reparametrization} to approximately bound $\calP$ in the Birkhoff polytope and 2) \emph{augmented Lagrangian method} (ALM) to push $\calP$ to a real permutation.
Hence we enable differentiable permutation optimization during the \emph{\supermesh weight} training phase.
We add an extra ALM term $\calL_{P}$ in the objective,
\begin{equation}
\small
\label{eq:ALMFormulation}
\begin{aligned}
    &\calL_P=\sum_{b=1}^{B_{max}}\sum_{i=1}^K\lambda^r_{b,i}\Delta\widetilde{\calP}_{b}^{i,:}+\!\sum_{b=1}^{B_{max}}\sum_{j=1}^K\lambda^c_{b,j}\Delta\widetilde{\calP}_{b}^{:,j}\\
    &+\frac{\rho}{2}\sum_{b=1}^{B_{max}}\sum_{i=1}^{K}\lambda^r_{b,i}(\Delta\widetilde{\calP}_b^{i,:})^2\!+\!\frac{\rho}{2}\sum_{b=1}^{B_{max}}\sum_{j=1}^{K}\lambda^c_{b,j}(\Delta\widetilde{\calP}_b^{:,j})^2,
\end{aligned}
\end{equation}
where $\lambda^r, \lambda^c\in\mathbb{R}^{B_{max}\times K}$ are the row-wise and column-wise Lagrangian multipliers, $\rho$ is the scalar quadratic penalty coefficient, and $\Delta$ denotes the difference between the $\ell_1$ norm and $\ell_2$ norm of the vector, e.g., $\Delta\widetilde{\calP}_b^{i,:}=\|\widetilde{\calP}_b^{i,:}\|_1-\|\widetilde{\calP}_b^{i,:}\|_2$.
This is different from the standard ALM formulation as the quadratic term is also controlled by $\lambda$.
In this way, the optimization is dominated by the task-specific loss at the beginning and gradually honors the constraint.

We \emph{reparametrize} $\calP_b$ as $\widetilde{\calP}_b$ to simplify the constraints in Eq.~\eqref{eq:PermutationLayerRelax}.
We (1) first apply absolute operation to the relaxed matrix to guarantee non-negativity, (2) then apply column-/row-wise normalization, and (3) finally apply row-wise soft projection to force binarization,
\begin{equation}
\small
\label{eq:Reparametrization}
\begin{aligned}
    &\calP_b'=\frac{|\calP_b|}{\textbf{1}^T|\calP_b|},\quad\calP_b''=\frac{\calP_b'}{\calP_b'\textbf{1}},\quad \widetilde{\calP}_b=\Omega_P(\calP_b'')\\
    &\Omega_P(\calP_b^{''i,j})=\left\{
    \begin{aligned}
        \texttt{Round}(\calP_b^{''i,j}) &~~\text{if }\max(\calP_b^{''i,:})\geq 1-\epsilon,\\
        \calP_b^{''i,j} &~~\text{if }\max(\calP_b^{''i,:})< 1-\epsilon\\
    \end{aligned}
    \right.,
\end{aligned}
\end{equation}
where $\epsilon$ is the projection threshold, typically set to 0.05.
The soft projection stops gradients when $\widetilde{\calP}$ is very close to a real permutation, which is designed to avoid gradient instability issues caused by an overly large linear penalty term as $\lambda$ quickly increases.

At each iteration in the \emph{\supermesh weight} training phase, we first update the relaxed permutation matrices using gradient-based methods,
then we update the Lagrangian multipliers as follows,
\begin{equation}
\small
    \label{eq:DualUpdate}
    \begin{aligned}
    \lambda^r_{b,i}~+\!= \rho\big(\Delta\widetilde{\calP}_b^{i,:}+\frac{1}{2}(\Delta\widetilde{\calP}_b^{i,:})^2\big), \quad
    \lambda^c_{b,j}~+\!=\rho\big(\Delta\widetilde{\calP}_b^{:,j}+\frac{1}{2}(\Delta\widetilde{\calP}_b^{:,j})^2\big).\\
    \end{aligned}
\end{equation}

\noindent\textbf{Stabilize Optimization via Initialization and Normalization.}~
The relaxed $\widetilde{\calP}$ cannot guarantee orthogonality during optimization. 
Thus cascading multiple such matrices ruins the orthogonality of $U$ and $V$ and causes training difficulty due to statistical instability.
To mitigate it, we initialize $\calP$ with a smoothed identity, i.e., $\calP^0=I(\frac{1}{2}-\frac{1}{2K-2})+\frac{1}{2K-2}$, shown in Fig.~\ref{fig:PermOptimization}.
Note that initializing it with random permutations does not work since no gradients will flow back to zero entries.
In addition, we propose a second technique to solve this via row-wise and column-wise $\ell_2$ normalization on the constructed $U$ and $V$, respectively.
By doing this, the normalized $U$ and $V$ can approximate the properties of true unitaries.
This normalization takes no effects when $U$ and $V$ converge to real unitaries but helps to stabilize the matrix statistics.

\noindent\textbf{Scheduling Coefficient $\rho$}.~
$\rho$ determines the speed to increase $\lambda$.
A large $\rho$ quickly traps $\widetilde{\calP}$ to a nearby suboptimal permutation.
An overly small $\rho$ has too weak constraints on the permutation.
Thus we increase $\rho$ as $\rho\gets\rho \gamma^t, \forall t\!=\!0,\cdots,T$, such that $\rho^T\approx 1e4\cdot\rho^0$.

\noindent\textbf{Stochastic Permutation Legalization (\texttt{SPL}).}~
Due to high non-convexity in the problem Eq.~\eqref{eq:ALMFormulation}, our ALM-based method does not guarantee convergence to a legal permutation.
Instead, it may stuck at saddle points shown in Fig.~\ref{fig:PermOptimization}.
To force $\widetilde{\calP}$ to a legal permutation after \supermesh training, we propose the following stochastic permutation legalization (\texttt{SPL}),
\begin{equation}
    \small
    \label{eq:PermutationLeaglization}
    \begin{aligned}
     PSQ^{*}=\texttt{SVD}\big(\texttt{Softmax}(\calP/\tau)\big)\big|_{\tau\rightarrow 0^+},~~
     \calP_{\texttt{SPL}}=\texttt{Softmax}\big((|PQ^{\ast}|+\delta)/\tau\big)\big|_{\tau\rightarrow 0^+},
    \end{aligned}
\end{equation}
where $\delta\sim\calN(0,\sigma^2)$.
We given an example in Fig.~\ref{fig:PermOptimization}.
The first \texttt{Softmax} binarizes the matrix.
Then, the SVD-based projection pushes the solution away from saddle points.
After that, random perturbations are added to break the ties between different rows.
The final \texttt{Softmax} pushes it into a legal permutation in a stochastic manner.
We repeat the second equation by multiple times until we find a legal solution without introducing too many extra crossings.

\subsubsection{Optimize Directional Couplers $\calT$}~
\label{sec:OptimizeCoupler}
The transmission coefficient $t$ of each directional coupler in the \dc layer is a binary optimization variable $t\in\{\frac{\sqrt{2}}{2},1\}$.
$t$=1 represents direct waveguide connection without placing a \dc.
We treat $t$ as a \supermesh \emph{weight} and perform quantization-aware training to learn the \dc layers.
The \dc binarization and its gradient are given as follows,
\begin{equation}
    \small
    \label{eq:QuantizedCoupler}
    \begin{aligned}
    T(t_q)&=T(\calQ(t)),~\calQ(t)=(\texttt{sign}(t)+1)\times\frac{2-\sqrt{2}}{4}+\frac{\sqrt{2}}{2},\\
    \frac{\partial\calL}{\partial t}&=\min\Big(1,\max\big(-1,\frac{\partial\calL}{\partial t_q}\times\frac{2-\sqrt{2}}{4}\big)\Big).
    \end{aligned}
\end{equation}

\subsubsection{Optimize Diagonal Matrix $\Sigma$ and Phases $\Phi$}~
\label{sec:OptimizeSigma}
We treat the diagonal matrix $\Sigma$ and phase shifter configurations $\Phi$ as the \supermesh \emph{weights}.
During \supermesh training, we simply apply the standard backpropagation to train them.

\begin{table*}[]
\centering
\caption{Evaluate searched PTCs with different sizes and footprint targets on MNIST with a 2-layer CNN.
The total block number is \#Blk=$B^U+B^V$.
\#\ps is omitted since we have \#\ps=$K\cdot$\#Blk.
All footprint constraints follow $F_{min}=0.8F_{max}$.
\name-a1 to \name-a5 cover 5 different footprint targets with the device specification from AMF foundry PDKs.
In the AMF PDKs~\cite{NP_AMF}, the footprint of \ps, \dc, and \crs is 6800 $\mu m^2$, 1500 $\mu m^2$, and 64 $\mu m^2$, respectively.
All footprint is reported in the unit of 1/1000 $\mu m^2$.
}
\vspace{-10pt}
\label{tab:CompareBlockSize}
\resizebox{0.9\textwidth}{!}{%
\begin{tabular}{cc|cc|ccccc}
\toprule
PTC Size               & Metrics         & MZI-ONN~\cite{NP_NATURE2017_Shen}    & FFT-ONN~\cite{NP_ASPDAC2020_Gu}    & \name-a1 & \name-a2 & \name-a3  & \name-a4   & \name-a5   \\ \midrule
\multirow{4}{*}{8$\times$8}   & \#\crs/\#\dc/\#Blk & 0/112/32   & 16/24/6    & 24/17/5  & 17/19/6  & 26/27/8   & 27/36/11   & 33/41/13   \\
                       & {[}$F_{min}$, $F_{max}${]} & -          & -          & {[}240, 300{]}  & {[}336, 420{]}   & {[}432, 540{]}   & {[}528, 660{]}   & {[}624, 780{]}   \\
                       & Footprint $\mathcal{F}$       & 1909       & 363        & 299      & 356      & 478       & 654        & 771        \\
                       & Accuracy (\%)   & 98.63      & 98.43      & 98.26    & 98.49    & 98.56     & 98.48      & 98.69      \\ \midrule
\multirow{4}{*}{16$\times$16} & \#\crs/\#\dc/\#Blk & 0/480/64   & 88/64/8    & 45/28/4  & 68/43/6  & 127/59/8  & 174/71/10  & 131/85/12  \\
                       & {[}$F_{min}$, $F_{max}${]} & -          & -          & {[}480, 600{]}  & {[}672, 840{]}   & {[}864, 1080{]}  & {[}1056, 1320{]} & {[}1248, 1560{]} \\
                       & Footprint $\mathcal{F}$      & 7683       & 972        & 480      & 722      & 967       & 1206       & 1441       \\
                       & Accuracy (\%)   & 98.65      & 98.25      & 98.16    & 98.40     & 98.24     & 98.56      & 98.57      \\ \midrule
\multirow{4}{*}{32$\times$32} & \#\crs/\#\dc/\#Blk & 0/1984/128 & 416/160/10 & 223/60/4 & 333/87/6 & 628/178/8 & 691/150/10 & 717/179/12 \\
                       & {[}$F_{min}$, $F_{max}${]} & -          & -          & {[}960, 1200{]} & {[}1344, 1680{]} & {[}1728, 2160{]} & {[}2112, 2640{]} & {[}2496, 3120{]} \\
                       & Footprint $\mathcal{F}$      & 30829      & 2443       & 975      & 1457     & 1959      & 2445       & 2926       \\
                       & Accuracy (\%)   & 98.68      & 97.97      & 98.10     & 98.18    & 98.36     & 98.49      & 98.39      \\ \bottomrule
\end{tabular}%
}
\vspace{-5pt}
\end{table*}

\subsection{PDK-Adaptive Footprint-Constrained \supermesh Optimization}
\label{sec:FootprintConstraint}
An important hardware constraint we need to honor is the target photonic circuit footprint, given the component sizes from a foundry PDK.
We solve the inequality footprint constraint by adding a \emph{probabilistic footprint penalty} term $\calL_{\calF}$,
\begin{equation}
    \small
    \label{eq:FootprintPenalty}
    \begin{aligned}
    &\calL_{\calF}=\left\{
    \begin{aligned}
    \beta\Big(\mathbb{E}[\calF_{\text{prox}}(\alpha)]/\widehat{F}_{max}\Big),& \quad \mathbb{E}[\calF(\alpha)]>\widehat{F}_{max},\\
    -\beta\Big(\mathbb{E}[\calF_{\text{prox}}(\alpha)]/\widehat{F}_{min}\Big),& \quad \mathbb{E}[\calF(\alpha)]< \widehat{F}_{min},\\
    0,& \quad \text{otherwise},
    \end{aligned}
    \right.\\
    &\mathbb{E}[\calF(\alpha)]=\sum_{b=1}^{B_{max}}m_{b,2}\calF_b,\quad \mathbb{E}[\calF_{\text{prox}}(\alpha)]=\sum_{b=1}^{B_{max}}m_{b,2}\calF_{b,\text{prox}},\\
    &\calF_b=\#\ps(\calR_{b})\cdot\calF_{\ps}+\#\dc(\calT_b)\cdot\calF_{\dc}+\#\crs(\calP_b)\cdot\calF_{\crs},\\
    &\calF_{b,\text{prox}}=\#\ps(\calR_{b})\cdot\calF_{\ps}+\#\dc(\calT_b)\cdot\calF_{\dc}+\beta_{\crs}\|\widetilde{\calP}_b-I\|_2^2\cdot\calF_{\crs},\\
    &\#\ps(\calR_b)=K,~\#\dc(\calT_b)=\sum_{i=1}^{(K-s_b)/2}\Big(\frac{2\calQ(t_i)}{\sqrt{2}-2}+\frac{2}{2-\sqrt{2}}\Big),
    \end{aligned}
\end{equation}
where $\beta$ is the penalty weight, and $\widehat{F}_{max}$ and $\widehat{F}_{min}$ is set to 0.95$F_{max}$ and 1.05$F_{min}$ to leave a 5\% constraint margin.
This penalty term allows \supermesh to control its \emph{expected footprint}.
Now we give a detailed breakdown of our probabilistic footprint penalty.

\noindent\textbf{Footprint of PS.}~
As an active device, \ps is not fixed after manufacturing.
Instead, the phase shifts $\Phi$ are important weights to guarantee enough PTC reprogrammability and ONN expressiveness.
Hence, we always assume a \emph{full column} of \ps, i.e., $\#\ps(\calR_b)=K$.

\noindent\textbf{Footprint of \dc.}~
\dc is typically fixed and not tunable. 
Hence the position to place a \dc need to be determined during the PTC design stage.
The footprint of a \dc layer is a simple summation of all couplers parameterized by their binarized coefficient $t_q$, which is fully differentiable by using straight-through estimators.

\noindent\textbf{Footprint of \crs.}~
The number of waveguide crossings, i.e., \#\crs, can be obtained by sorting rows of the permutation $\calP_b$ to an identity $I$ and finding the \emph{minimum number of adjacent swaps}.
However, this crossing counting procedure \#\crs($\calP_b$) itself is non-differentiable.
When calculating the footprint penalty, we replace $\#\crs(\calP_b)\calF_{\crs}$ with a differentiable proxy term $\beta_{\crs}\calF_{\crs}\|\widetilde{\calP_b}-I\|_2^2$, where $\beta_{\crs}$ is used to balance the penalty on \dc and \crs.

\noindent\textbf{Analytical Bound of the \supermesh Block Number.}~
Given the device footprint specification, we can actually calculate the maximum/minimum footprint of each block.
Based on the target footprint, we can find an analytical bound of the block number for our \supermesh, i.e., $B_{max}$ and $B_{min}$, without manual definition,
\begin{equation}
    \small
    \label{eq:BlockBound}
    \begin{aligned}
    \calF_{b,min}&\!=\!K\calF_{\ps}+\calF_{\dc}, ~~
    \calF_{b,max}\!=\!\calF_{b,min}+K\calF_{\dc}/2+K(K\!-\!1)\calF_{\crs}/2,\\
    B_{max}&=\ceil{F_{max}/\calF_{b,min}},\quad B_{min}=\floor{F_{min}/\calF_{b,max}}.
    \end{aligned}
\end{equation}

\section{Experimental Results}
\label{sec:ExperimentalResults}
\subsection{Experiment Setup}
\label{sec:ExpSetup}
\noindent\textbf{Datasets.} We search PTCs on MNIST and evaluate on MNIST, FashionMNIST,
SVHN~\cite{NN_svhn2011}, and CIFAR-10 datasets.

\noindent\textbf{NN Models.}~
We perform \supermesh training on MNIST with a 2-layer CNN (C32K5-BN-ReLU-C32K5-BN-ReLU-Pool5-FC10), where C32K5 is a 32-channel convolution with a kernel size of 5$\times$5.
In variation-aware training, we use LeNet-5 and VGG-8.

\noindent\textbf{Training Settings.}~
We train \supermesh for 90 epochs using Adam optimizer with an initial learning rate (lr) of 0.001 and a cosine lr scheduler.
We set the weight decay rate to 1e-4 for $\Phi$ and $\Sigma$, and 5e-4 for $\theta$.
We exponentially decrease the Gumbel-softmax temperature $\tau$ from 5 to 0.5.
We set 10 epochs in the \supermesh \emph{Warmup} stage. 
In the \supermesh \emph{Search} stage, we train weights and arch. params with a ratio of 3:1.
In the permutation ALM, we set the initial $\rho^0$=(1e-7)$\times K$/8.
We set $\beta$ and $\beta_{\crs}$ to 10 and 100 in the footprint penalty.
At the 50-th epoch, we force $\calP$ to a legal permutation by stochastic permutation legalization (\texttt{SPL}).
Then we continue the alternate \supermesh training in the rest 40 epochs.
During re-training, we sample a \submesh from the learned distribution $P_{\theta}$ that satisfies the footprint constraints.
Then we perform variation-aware training with Gaussian phase noises $\Delta\phi\sim\calN(0,0.02^2)$ to increase robustness.
\vspace{-20pt}
\subsection{Main Results}
\label{sec:MainResults}
We search PTC topologies with the proposed \name flow on three different PTC sizes (8$\times$8, 16$\times$16, 32$\times$32) with various footprint constraints.
We denote our searched PTC designs as \name-a1 to \name-a5.
In Table~\ref{tab:CompareBlockSize}, we compare our \name-series to prior manual PTC designs, i.e., MZI-ONN~\cite{NP_NATURE2017_Shen} and FFT-based ONN~\cite{NP_ASPDAC2020_Gu, NP_TCAD2020_Gu} on AMF foundry PDKs.
For a fair comparison, the butterfly mesh in the FFT-based PTC is not limited to Fourier-transform but a general trainable transform~\cite{NP_TCAD2020_Gu}.
On three PTC sizes, the searched \name-series shows superior adaptability to various footprint constraints.
Compared to the largest MZI-based PTC, our \name-series shows competitive learnability with \textbf{2$\times$-30$\times$} footprint reduction.
\name-series outperforms the FFT-based PTC with higher expressivity, especially on large PTC sizes, and saves up to \textbf{2.5$\times$} area.
\name shows superior adaptability to balance footprint and expressiveness.

\begin{table*}[]
\centering
\caption{MNIST accuracy with 16$\times$16 PTCs on AIM photonics PDKs~\cite{NP_OFC2018_Timurdogan}, where $\calF_{\ps}$=2500 $\mu m^2$, $\calF_{\dc}$=4000 $\mu m^2$, and $\calF_{\crs}$=4900 $\mu m^2$.
}
\vspace{-10pt}
\label{tab:CompareAIM}
\resizebox{0.95\textwidth}{!}{%
\begin{tabular}{cc|cc|cccccc}
\toprule
PTC Size               & Metrics          & MZI-ONN~\cite{NP_NATURE2017_Shen}  & FFT-ONN~\cite{NP_ASPDAC2020_Gu, NP_TCAD2020_Gu} & \name-a0       & \name-a1       & \name-a2       & \name-a3        & \name-a4         & \name-a5         \\ \midrule
\multirow{4}{*}{16$\times$16} & \#\crs/\#\dc/\#Blk  & 0/480/64 & 88/64/8 & 15/35/5        & 1/58/8         & 26/58/8        & 17/92/13        & 25/99/14         &  89/111/16                \\
                       & {[}$\mathcal{F}_{min}$, $\mathcal{F}_{max}${]} & -        & -       & {[}384, 480{]} & {[}480, 600{]} & {[}672, 840{]} & {[}864, 1080{]} & {[}1056, 1320{]} & {[}1248, 1560{]} \\
                       & Footprint $\mathcal{F}$        & 4480     & 1007    & 414            & 557            & 679            & 971             & 1079             &   1520               \\
                       & Accuracy (\%)    & 98.77    & 98.10   & 98.15          & 98.30          & 98.32          & 98.55           & 98.64            & 98.72                 \\ \bottomrule
\end{tabular}%
}
\vspace{-10pt}
\end{table*}

\noindent\textbf{Adapt PTCs to Different Foundry PDKs.}~To adapt \name to different device specifications, we change the foundry PDK from AMF~\cite{NP_AMF} to AIM photonics~\cite{NP_OFC2018_Timurdogan}, which provides much larger waveguide crossings.
In Table~\ref{tab:CompareAIM}, \name finds feasible PTC topology that avoids using many crossings to honor the strict footprint constraints.
The smallest \name-a0 achieves comparable accuracy to the FFT-based PTC with \textbf{2.4$\times$} smaller footprint.
Compared to MZI-based PTC, our \name-a5 is \textbf{2.9$\times$} more compact with similar expressiveness.

\begin{table}[]
\centering

\caption{Adapt searched 16$\times$16 PTCs to LeNet-5/VGG-8 and different datasets on AMF PDKs.
Test accuracy (\%) is given in the table.
The PTC is searched on MNIST and a 2-layer CNN.
}
\vspace{-5pt}
\label{tab:CompareDataset}
\resizebox{0.9\columnwidth}{!}{%
\begin{tabular}{cccc|cc}
\toprule
Model                    & \multicolumn{1}{c|}{Datasets} & MZI~\cite{NP_NATURE2017_Shen} & FFT~\cite{NP_ASPDAC2020_Gu, NP_TCAD2020_Gu} & \name-a2 & \name-a4 \\ \midrule
\multicolumn{2}{c|}{Footprint}                            & 7683    & 972     & 722      & 1206     \\ \midrule
\multirow{3}{*}{LeNet-5} & \multicolumn{1}{c|}{FMNIST}   & 87.33   & 85.87   & 85.89    & 87.07    \\
                         & \multicolumn{1}{c|}{SVHN}     & 69.91   & 65.04   & 65.26    & 69.20    \\
                         & \multicolumn{1}{c|}{CIFAR-10} & 51.40   & 42.75   & 51.26    & 52.42    \\ \midrule
\multirow{3}{*}{VGG-8}   & \multicolumn{1}{c|}{FMNIST}   & 89.59   & 88.62   & 89.23    & 89.16    \\
                         & \multicolumn{1}{c|}{SVHN}     & 77.87   & 75.22   & 75.86    & 77.20    \\
                         & \multicolumn{1}{c|}{CIFAR-10} & 68.90   & 63.57   & 66.30    & 68.50    \\ \bottomrule
\end{tabular}%
}
\vspace{-5pt}
\end{table}

\noindent\textbf{Transfer to Different ONNs and Datasets.}~
\label{sec:CompareDataset}
To further validate the expressiveness of \name-series searched on a proxy NN model and dataset, we apply searched PTC structures to other NN architectures and more challenging datasets in Table~\ref{tab:CompareDataset}.
On three datasets with LeNet-5 and VGG-8, our searched 16$\times$16 \name-a2 and \name-a4 significantly outperform FFT-based design with much higher accuracy and 26\% footprint reduction.
Compared to the MZI-based PTC, \name-a4 can save over \textbf{84\%} footprint with competitive performance.

\noindent\textbf{Noise Robustness of Searched PTCs.}~
\label{sec:Robustness}
\begin{figure}
    \centering
    \vspace{-10pt}
    \subfloat[]{\includegraphics[width=0.42\columnwidth]{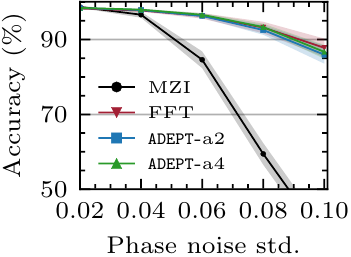}
    \label{fig:RobustnessMNIST}
    }
    \subfloat[]{\includegraphics[width=0.42\columnwidth]{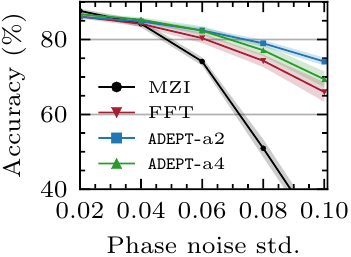}
    \label{fig:RobustnessFMNIST}
    }
    \vspace{-10pt}
    \caption{Robustness evaluation of 16$\times$16 PTCs with various phase noise intensities.
    (a) 2-layer CNN on MNIST.
    (b) LeNet-5 on FMNIST.
    All models are trained with variation-aware training.
    The shadow marks $\pm 3\sigma$ uncertainty over 20 runs.
    }
    \label{fig:CompareRobustness}
\end{figure}

In Fig.~\ref{fig:CompareRobustness}, we inject phase drifts into the circuit and perform variation-aware training on all PTC designs~\cite{NP_ICCAD2019_Zhao, NP_DATE2020_Gu}.
Even with noise-aware training, the MZI-based ONN still suffers a severe accuracy drop due to overly large PTC depth.
In contrast, our searched PTCs show similar or even better noise robustness than the logarithmic-depth FFT-based design.

\subsection{Ablation Studies}
\label{sec:PermutationAblation}

\begin{figure}
    \centering
    \vspace{-15pt}
    \subfloat[]{\includegraphics[width=0.24\textwidth]{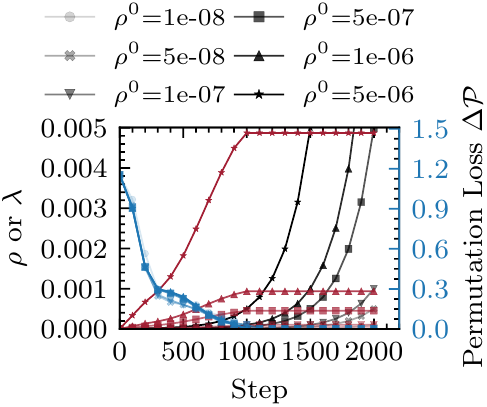}
    \label{fig:PermCurve}
    }
    \subfloat[]{\includegraphics[width=0.24\textwidth]{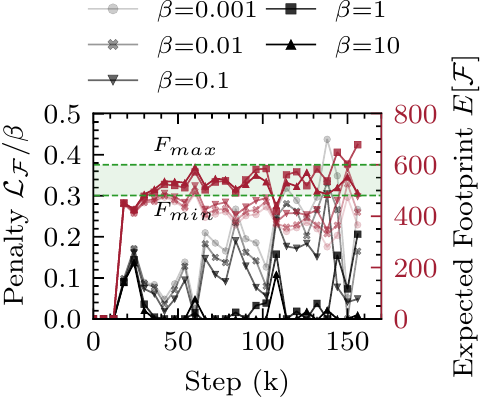}
    \label{fig:FootprintCurve}
    }
    \vspace{-13pt}
    \caption{(a) Scan initial $\rho$ in permutation ALM from 5e-8 to 5e-6.
    Red lines are averaged $\lambda$.
    Blue curves are permutation errors, i.e., average difference between $\ell_1$- and $\ell_2$-norm.
    (b) Scan $\beta$ in footprint penalty from 0.001 to 10.
    Red lines are expected footprint $\mathbb{E}[\mathcal{F}(\alpha)]$ of \name-a1.
    Black curves are footprint penalty.
    The green region marks the constraint.
    }
    \vspace{-10pt}
\end{figure}

\noindent\textbf{Permutation ALM.}~To better understand the permutation learning process, we scan different initial values of the ALM penalty coefficient $\rho^0$ and plot the optimization curves in Fig.~\ref{fig:PermCurve}.
Our method is insensitive to the hyper-parameter settings and can stably converge with the proposed adaptive penalty scheduling.

\noindent\textbf{Footprint Penalty.}~
\label{sec:FootprintAblation}
In Fig.~\ref{fig:FootprintCurve}, the expected PTC footprint is visualized with different penalty strengths.
With $\beta$=$\sim$10, the expected footprint of \supermesh can be well-bounded.
If $\beta$ is too small, most sampled PTC structures from $P_{\theta}$ will violate the constraint.

\section{Conclusion}
\label{sec:Conclusion}
In this work, \emph{for the first time}, we propose an automatic differentiable framework \name for efficient photonic tensor core design.
Our \name constructs a probabilistic photonic \supermesh, employs an augmented Lagrangian method to learn waveguide connections, and adopts binarization-aware training to search coupler locations.
With a probabilistic footprint penalty method, \name integrates circuit area constraints into \supermesh training procedure to adapt the PTC to various device specifications and footprint constraints.
Extensive experiments show the superior flexibility of \name for automated PTC topology search adaptive to foundry PDKs.
The searched PTC design outperforms prior manual designs with competitive expressiveness, \textbf{2$\times$-30$\times$} smaller footprint, and superior robustness.
\name opens a new paradigm in photonic neurocomputing by "nurturing" photonic circuit design via AI and automation.

\vspace{-5pt}
\begin{acks}
The authors acknowledge the Multidisciplinary University Research Initiative (MURI) program through the Air Force Office of Scientific Research (AFOSR), contract No. FA 9550-17-1-0071, monitored by Dr. Gernot S. Pomrenke.
\end{acks}


\end{document}